\documentclass[11pt]{article}

\usepackage{amssymb,graphicx,fullpage}

\def\Arctan{\mathop{\textrm{Arctan}}\nolimits}

\newcommand{\pard}[2]{\frac{\partial {#1}}{\partial {#2}}}

\begin{document}

\title{Control of spatio-temporal patterns in the Gray-Scott model}

\author{Y.N. Kyrychko\thanks{Corresponding author. Email: y.kyrychko@sussex.ac.uk},\hspace{0.2cm} K.B. Blyuss
\\\\Department of Mathematics, University of Sussex,\\Brighton, BN1 9QH, United Kingdom\\ \vspace{0.2cm}
\and S.J. Hogan\\\\
Department of Engineering Mathematics, University of Bristol,\\ Bristol, BS8 1TR, United Kingdom\\ \vspace{0.2cm}
\and E. Sch\"oll\\\\
Institut f\"ur Theoretische Physik, Technische Universit\"at Berlin,\\ 10623 Berlin, Germany}

%

\maketitle

\begin{abstract}
The paper studies the effects of a time-delayed feedback control on the appearance and development of spatio-temporal patterns in a reaction-diffusion system. Different types of control schemes are investigated, including single-species, diagonal and mixed control. This approach helps to unveil different dynamical regimes, which arise from chaotic state or from tavelling waves. In the case of spatio-temporal chaos, the control can either stabilize uniform steady states or lead to bistability between a trivial steady state and a propagating travelling wave. Furthermore, when the basic state is a stable travelling pulse, the control is able to advance stationary Turing patterns, or yield the above-mentioned bistability regime. In each case, the stability boundary is found in the parameter space of the control strength and the time delay, and numerical simulations suggest that diagonal control fails to control the spatio-temporal chaos.
\end{abstract}

{\bf Various real-life systems exhibit complex dynamical regimes, including spatial patterns and spatio-temporal chaos. An important issue is the possibility of attaining a desired state by means of some external influence on the system. Time-delayed feedback uses the difference between current state of the system and its state some time ago to provide an efficient  tool to control system's dynamics. In particular, this approach allows one to effectively switch between different spatio-temporal regimes in the system. In this paper, we show how the time-delayed feedback control can be used to control spatio-temporal chaos, as well as to provide a transition from travelling waves to stationary periodic patterns.
}

\section{Introduction}

Spatially extended systems arise in modelling of various physical, biological, chemical and engineering phenomena, and an ultimate understanding of such systems gives one the ability to control them in order to achieve desired behaviour or spatial patterns. 
Since many systems in nature exhibit intrinsic chaotic behaviour, control of chaos in spatially extended systems has recently become 
an active and important area of research.

Control of a dynamical system in a chaotic state usually requires application of some external perturbation in order to achieve a desired 
type of behaviour, such as a steady state, regular or quasi-periodic oscillations. A number of different methods and techniques have been 
suggested in order to control chaos \cite{SH}. For systems without spatial extension, standard approaches include the Ott-Grebogi-Yorke (OGY) 
scheme \cite{ogy} and time-delayed feedback control (TDFC) \cite{pyr}. TDFC schemes are straightforward to implement and can also be adapted 
to control spatially extended systems. These schemes are based on the use of the difference between the system variables at the current moment of time 
and their values at some time in the past. For low-dimensional systems, TDFC has been successfully used to stabilize unstable steady 
states \cite{HS,DHS,SHWSH}, as well as unstable periodic orbits \cite{FIE07} in a number of physical and biological systems \cite{SH}.

More recently, TDFC has been used to control spatio-temporal chaos and pattern formation in chemical reactions. Close to the onset of uniform oscillations, it is possible to reduce any reaction-diffusion system to an amplitude equation, called the complex Ginzburg-Landau equation (CGLE) \cite{Ku}. The CGLE helps to understand the underlying dynamics of the reaction-diffusion systems  close to the bifurcation point, however, it does not provide a lot of information about the dynamics far away from the onset of oscillations. Beta and Mikhailov have shown analytically that global TDFC cannot stabilize uniform oscillations when the system in the spatio-temporally chaotic regime \cite{Beta2,KIM01}. At the same time, local time-delay feedback control which also includes spatial feedback terms has been shown to be effective in stabilizing travelling waves in CGLE in one and two spatial dimensions \cite{MoS,PS}.

In globally coupled reaction-diffusion systems, which arise in modelling semiconductor nanostructures, global or local TDFC and its various modifications have proved successful in stabilizing unstable spatio-temporal oscillations \cite{BASSJ,FRA99,UAJS}. Unstable rigid rotation of spiral waves has also been stabilized by TDFC \cite{SCH06}. Pulse propagation in the FitzHugh-Nagumo reaction-diffusion model has recently been shown to be controllable by local TDFC or nonlocal instantaneous feedback \cite{DAH08}.

From a practical perspective, there are several issues that have to be taken into account when considering implementation of specific TDFC schemes. One such aspect concerns the spatial range over which the control has to be applied. Ahlborn and Parlitz \cite{AP} have recently shown how one can use multiple time-delayed control applied at a number of control points to stabilize plane waves in the CGLE, while Bleich and Socolar \cite{BS} have illustrated stabilization of periodic orbits in chaotic CGLE using local control. Another problem is that most examples of TDFC for spatially-extended systems such as, for instance, one-dimensional chaotic CGLE \cite{BS} and spatio-temporally chaotic semiconductor laser arrays \cite{MKH}, use diagonal control with symmetric gain, which may be quite restrictive. In the case of optical systems such as laser devices it is possible to have a non-diagonal coupling, realized experimentally using an optical phase as an additional control parameter \cite{DHS,SHWSH}. Hence, it is important to consider the influence of various TDFC schemes  on the dynamics of the system and analyze different types of spatio-temporal dynamical regimes that can be achieved.

In this paper we perform a systematic study of the effects of local TDFC on spatio-temporal dynamics in a two-component reaction-diffusion 
system. As a testbed for this analysis we use the paradigmatic Gray-Scott model derived in the context of chemical reactions. It has already been shown using extensive numerical simulations that the Gray-Scott model supports a wide range of spatio-temporal dynamics, including stationary inhomogeneous patterns, travelling fronts and pulses, as well as spatio-temporal chaos. A particular feature of this system, which makes it different from other reaction-diffusion models, is the occurrence of pulse-splitting, where a travelling pulse leaves in its wake other pulses propagating in different directions. 

The outline of this paper is as follows. In the next section we introduce the Gray-Scott system. Section 3 contains an analysis of the effects of TDFC on spatio-temporal chaos. In Section 4 various TDFC schemes are applied to propagating pulses. The paper concludes with a summary of our findings, together with a consideration of their implications.

\section{Gray-Scott system}

In order to analyse the effects of TDFC on the spatio-temporal dynamics in reaction-diffusion systems, we consider the Gray-Scott model, which describes a cubic auto-catalytic chemical reaction of the type \cite{GS}
\begin{equation}\label{eq1}
\begin{array}{l}
U+2V\to 3V,\\
V\to P,
\end{array}
\end{equation}
where $U$ is continuously supplied into an open flow reactor, and the product $P$ is removed. The first equation represents an autocatalytic process, in which two molecules of species $V$ produce $3V$ through interaction with a molecule of species $U$. If the system were closed, then by virtue of irreversibility of the two reactions, all reactants would eventually turn into a product and be removed. However, when the reactor is continuously fed with a uniform flow of species $U$, it is possible to maintain far-from-equilibrium conditions, in which a wide range of possible dynamics can be observed. In experiments, a gel is used as a medium to prevent the occurrence of convective currents \cite{LMOS}. The kinetic equations for the system (\ref{eq1}) can be written after some rescaling as
\begin{equation}\label{GS1}
\begin{array}{l}
\displaystyle{\pard{u}{t}=-uv^{2}+a(1-u)+D_{u}\nabla^{2}u=f(u,v)+D_{u}\nabla^{2}u,}\\\\
\displaystyle{\pard{v}{t}=uv^{2}-(a+b)v+D_{v}\nabla^{2}v=g(u,v)+D_{v}\nabla^{2}v,}
\end{array}
\end{equation}
where $u$ and $v$ are the concentrations of the species $U$ and $V$, respectively, $a$ is the inflow rate, $a+b$ is the removal rate of $V$ from the reaction, and $D_{u}$ and $D_{v}$ are the diffusion coefficients of the two species.

The system (\ref{GS1}) can possess up to three homogeneous steady states. There is a trivial steady state
\begin{equation}
E_{0}=(u_{0},v_{0})=(1,0),
\end{equation}
which exists for all parameter values and corresponds to a steady flow without reaction. Provided
\begin{equation}
d\equiv 1-4(a+b)^{2}/a>0,
\end{equation}
the system has two non-trivial steady states
\begin{equation}\label{SS1}
E_{1}=\displaystyle{(u_{1},v_{1})=\left(\frac{1}{2}(1-\sqrt{d}),\frac{1}{2}\frac{a}{a+b}(1+\sqrt{d})\right),}
\end{equation}
and
\begin{equation}
E_{2}=\displaystyle{(u_{2},v_{2})=\left(\frac{1}{2}(1+\sqrt{d}),\frac{1}{2}\frac{a}{a+b}(1-\sqrt{d})\right).}
\end{equation}

\begin{figure*}
\hspace{-1cm}\includegraphics[width=18cm]{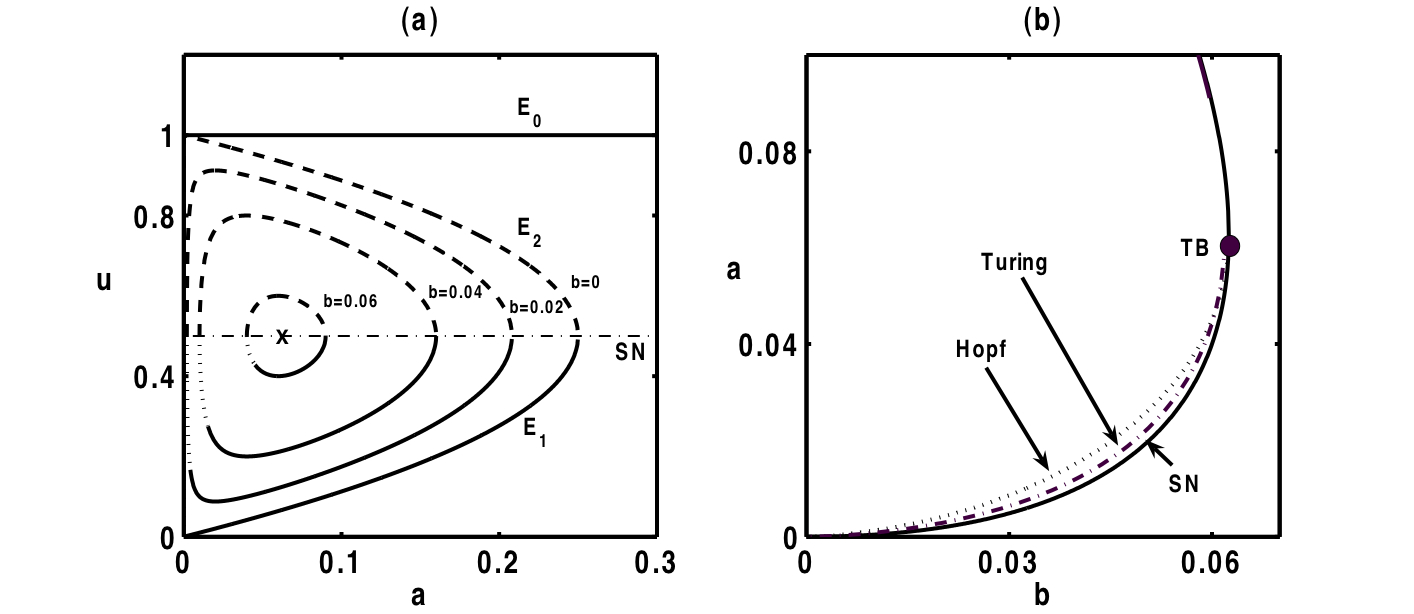}
\caption{(a) Bifurcation diagram of the Gray-Scott system (\ref{GS1}) without diffusion. Solid lines correspond to stable steady states, dashed to the unstable steady state $E_{2}$, dotted to the Hopf unstable steady state $E_{1}$, and dash-dotted indicates a saddle-node (SN) line. The cross (X) indicates the location of the Takens-Bogdanov point. (b) Phase diagram in $(a,b)$ space. Solid line is the boundary of saddle-node bifurcation, dotted line is the boundary of Hopf bifurcation, and dash-dotted line is Turing instability boundary. TB is the Takens-Bogdanov point.}\label{SN}
\end{figure*}

Linear stability analysis indicates that the trivial steady state $E_{0}$ is always linearly stable (even with respect to spatially inhomogeneous perturbations), while the steady state $E_{2}$ is unstable with respect to homogeneous perturbations for any parameter values for which it exists. Figure~\ref{SN} shows a bifurcation diagram for the spatially homogeneous Gray-Scott model, as well as a phase diagram in $(a,b)$ space. One of the features is that when the inflow $a$ is large enough ($a>1/4$), the production of $V$ is suppressed, and as a result the only fixed point of the system is the trivial steady state $E_{0}$, as shown in Fig.~\ref{SN}(a). 

In Fig.~\ref{SN}(b) the saddle-node bifurcation curve is given by
\begin{equation}
\displaystyle{a^{\rm SN}=\frac{1}{8}[1-8b\pm\sqrt{1-16b}]}.
\end{equation}
To the right of  $a^{\rm SN}$ only the trivial steady state $E_{0}$ exists. As $a^{\rm SN}$ is crossed, two more states $E_{1}$ and $E_{2}$ appear. While the state $E_{2}$ is always unstable, the fixed point $E_{1}$ may lose its stability via a Hopf bifurcation. The boundary of this bifurcation is given by \cite{MRMBD}
\begin{equation}
\displaystyle{a^{\rm H}=\frac{1}{2}\left[\sqrt{b}-2b-\sqrt{b(1-4\sqrt{b})}\right]}.
\end{equation}
This boundary starts at $(a,b)=(0,0)$ and follows the lower branch of the saddle-node curve until it coincides with it at a Takens-Bogdanov point $(a,b)=(1/16,1/16)$. The steady state $E_{1}$ is stable above the Hopf boundary and unstable below it. It is noteworthy that the stability of the limit cycle that appears due to the Hopf bifurcation varies along the Hopf curve: it is stable for $b<0.035$ and unstable for $b>0.035$. 

Another way in which a state $E_{1}$ may lose its stability is via a Turing bifurcation, which may give rise to stationary inhomogeneous patterns. Linear stability analysis in this case shows that the Turing bifurcation occurs when the ratio of the two diffusion coefficients $\sigma=D_{u}/D_{v}$ satisfies the condition \cite{MRMBD}
\begin{equation}
\left[\sigma(a+b)-(v_{1}^{2}+a)\right]^{2}>4\sigma(a+b)(v_{1}^{2}-a),
\end{equation}
where $v_{1}$ is the steady-state value of the species $V$ given in equation (\ref{SS1}). The higher $\sigma$, the wider is the region in $(a,b)$ space where Turing bifurcation can be observed. The steady state $E_{1}$ is stable with respect to spatially inhomogeneous perturbations above the Turing boundary and unstable below it. For sufficiently high values of $\sigma$, the boundary of Turing bifurcation will move above the boundary of Hopf bifurcation. For $\sigma=2$, which will be used in numerical simulations later in this paper, the region of Turing instability is almost entirely contained within the parameter region of Hopf instability, as shown in Fig.~\ref{SN}(b). Interaction of these two instabilities gives rise to spatio-temporal chaos and mixed modes having time-dependent spatial structures \cite{MRMBD,NU}. In Figure~\ref{SN}(b), the region of spatio-temporal chaos is narrow and bounded by the Turing boundary from the top and the saddle-node curve from the bottom. It is noteworthy that very close to the Takens-Bogdanov point one has a Turing instability where stationary inhomogeneous patterns are observed.

It has already been mentioned that the Gray-Scott system exhibits a wide range of dynamical behaviours. In the 
parameter regime away from Turing and Hopf areas, the system always ends up in either the homogeneous trivial steady state $E_{0}$ or state $E_{1}$. The simplest inhomogeneous state is represented by a stationary non-uniform pattern arising from a Turing bifurcation. Besides these temporally stationary structures, the Gray-Scott system is able to support standing or travelling waves and fronts (the latter refer to heteroclinic connections between $E_{0}$ and $E_{1}$). Travelling waves occur in the parameter region outside the saddle-node boundary. At a higher level of complexity are the mixed modes, describing time-dependent spatial structures, the so-called self-replicating patterns and spatio-temporal chaos \cite{NU}. The self-replicating patterns represent heteroclinic connections between the trivial homogeneous state and a spatially periodic pattern.

\section{Control of spatio-temporal chaos}

In this section we concentrate on the parameter regime where the system exhibits spatio-temporal chaos and consider several schemes 
for the control of this state based on TDFC. This approach has already been used to control spatio-temporal chaos in globally coupled 
reaction-diffusion systems \cite{BASSJ,UAJS}, as well as in reaction-diffusion systems with global control \cite{Beta2, MS}.

Formally, the locally controlled Gray-Scott model can be written as
\begin{equation}\label{GS}
\pard{}{t}\left(
\begin{array}{l}
u\\
v
\end{array}
\right)=
\left(
\begin{array}{l}
f(u,v)\\
g(u,v)
\end{array}
\right)+\left(
\begin{array}{ll}
D_{u}&0\\
0&D_{v}
\end{array}
\right)
\nabla^{2}\left(
\begin{array}{l}
u\\
v
\end{array}
\right)+KA\left(
\begin{array}{l}
u(t-\tau)-u(t)\\
v(t-\tau)-v(t)
\end{array}
\right),
\end{equation}
where $K$ is the control strength which can be either positive or negative, $\tau>0$ is the time delay, and $A$ is a two-by-two matrix which describes the particular coupling of the control term. 
First we consider the case when the control of each variable depends only on the history of that variable. For single-species control, the matrix $A$ takes the form 
\begin{equation}\label{ss_control}
A_{u}=\left(
\begin{array}{ll}
1&0\\
0&0
\end{array}
\right),\hspace{1cm}\mbox{or}\hspace{1cm}
A_{v}=\left(
\begin{array}{ll}
0&0\\
0&1
\end{array}
\right),
\end{equation}
for control only in $u$ (activator control) and only in $v$ (inhibitor control), respectively.

In the case of activator control, we look for solutions of the system (\ref{GS}) linearized near $E_{1}=(u_{1},v_{1})$ in the form
\[
\left(
\begin{array}{l}
\tilde{u}\\
\tilde{v}
\end{array}
\right)=\left(
\begin{array}{l}
C_{1}\\
C_{2}
\end{array}
\right)e^{iqx+\lambda t},
\]
where $u=u_{1}+\tilde{u}$, $v=v_{1}+\tilde{v}$, $C_{1,2}$ are constants, $q$ is the wavenumber of the perturbation, and $\lambda$ is the corresponding growth rate. Substituting this into the system (\ref{GS}) with $A_{u}$ from (\ref{ss_control}) yields the following characteristic equation
\begin{equation}\label{char_eq_mat}
\left|
\begin{array}{cc}
-v_{1}^{2}-a-D_{u}q^{2}+K[e^{-\lambda\tau}-1]-\lambda&-2(a+b)\\
v_{1}^{2}&a+b-D_{v}q^{2}-\lambda
\end{array}
\right|=0.
\end{equation}
\begin{figure*}
\hspace{-1.5cm}\includegraphics[width=18cm]{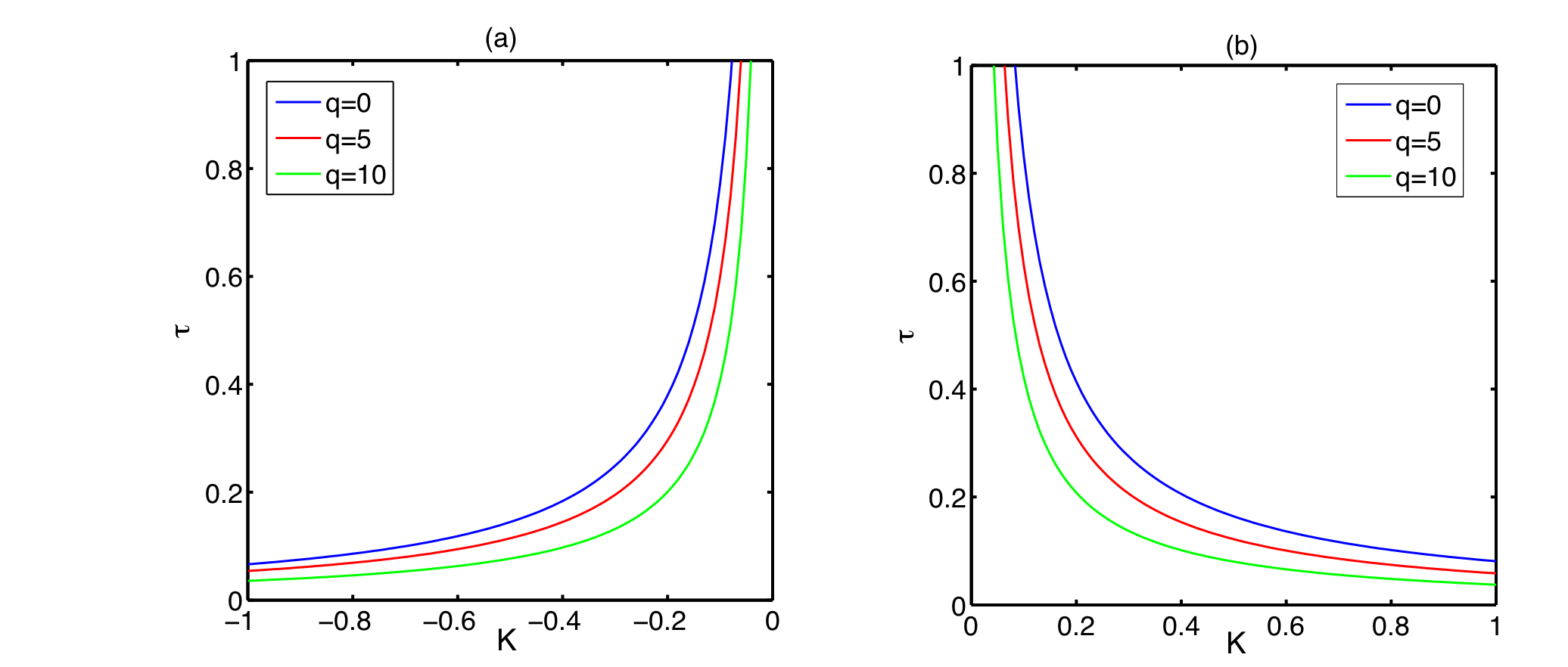}
\caption{(Color online) Stability boundaries of the homogeneous steady state $E_{1}$ from the characteristic equation (\ref{char_eq_mat}). In both cases, the steady state $E_{1}$ is stable above the boundary for $q=0$ described by (\ref{stability}). (a) Activator control. (b) Inhibitor control. Parameter values are $a=0.028$, $b=0.053$, $D_{u}=2\cdot 10^{-5}$, $D_{v}=10^{-5}$.}\label{stab_bounds}
\end{figure*}
In the absence of control $(K=0)$, it follows from this equation that the uniform mode with $q=0$ is the most unstable, i.e. it has an eigenvalue with the largest positive real part. Since the diffusion plays a stabilizing role, all other modes with $q>0$ will be more stable, as is confirmed by numerical solution of equation (\ref{char_eq_mat}) shown in Fig.~\ref{stab_bounds}. For this reason, it suffices to control the uniform mode $q=0$ to achieve complete stabilization of the steady state $E_{1}$ for any spatially inhomogeneous perturbation. When $q=0$, the equation (\ref{char_eq_mat}) can be written as
\begin{equation}\label{char_eq}
\lambda^{2}+\lambda\left[v_{1}^{2}-b-K\left(e^{-\lambda\tau}-1\right)\right]+(a+b)\left[v_{1}^{2}-a+K\left(e^{-\lambda\tau}-1\right)\right]=0.
\end{equation}
In order to find the control strength $K$ that can stabilize the steady state $E_{1}$ for each value of the time delay $\tau$, we compute the stability boundary in the $(K,\tau)$ plane by setting $\lambda=i\omega$ in equation (\ref{char_eq}). Separating real and imaginary parts gives the following system
\[
\begin{array}{l}
-\omega^{2}+(a+b)(v_{1}^{2}-a-K)-K\omega\sin\omega\tau+(a+b)K\cos\omega\tau=0,\\
\omega(-b+v_{1}^{2}+K)-\omega K\cos\omega\tau-(a+b)K\sin\omega\tau=0.
\end{array}
\]
This system can be solved to yield the pair $(K,\tau)$ as parameterized by $\omega$:
\begin{equation}\label{stability}
\begin{array}{l}
\displaystyle{K=-\frac{1}{2}\frac{\left[(a+b)^2+\omega^{2}\right]\left[(a-v_{1}^{2})^{2}+\omega^{2}\right]-4\omega^{2}bv_{1}^{2}}{(a-v_{1})^{2}(a+b)^{2}+\omega^{2}(a+v_{1}^{2})}},\\\\
\displaystyle{\tau=\frac{1}{\omega}\left[\Arctan\left(\frac{2\omega}{K}-\frac{4(a+b)v_{1}^{2}\omega}{K\left[(a+b)^{2}+\omega^{2}\right]}\right)\pm n\pi\right]},\hspace{0.5cm} n=0,1,2,..,
\end{array}
\end{equation}
\begin{figure*}
\hspace{-0.5cm}\includegraphics[width=17.5cm]{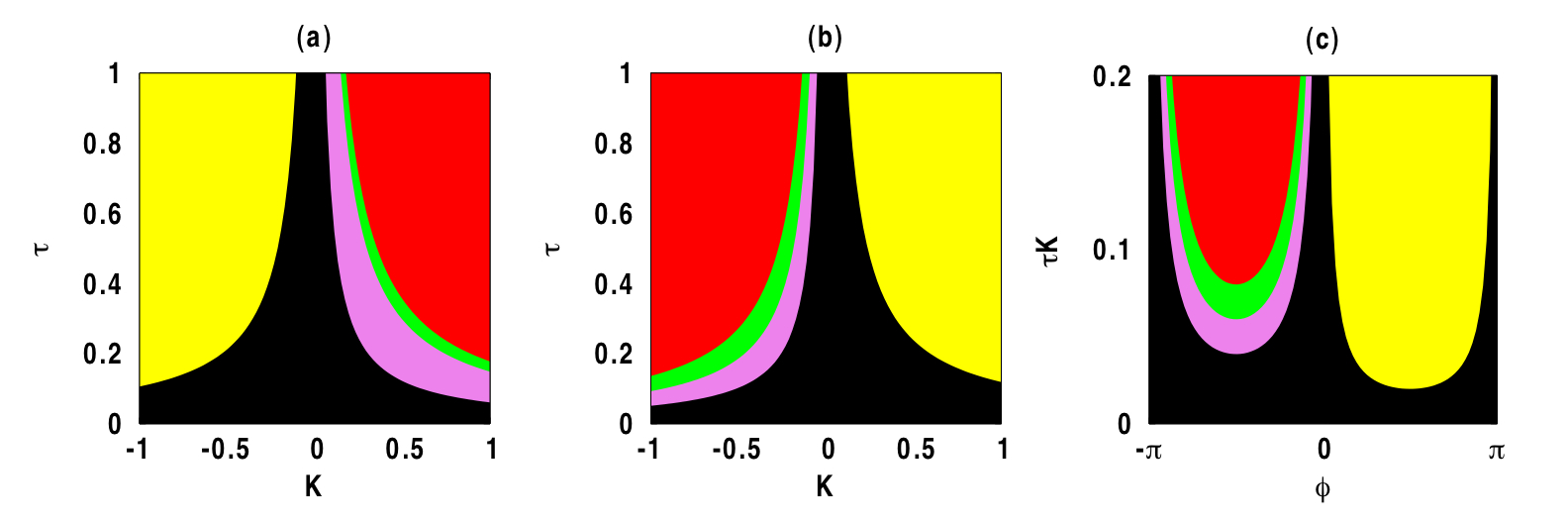}
\caption{(Color online) Control domain of spatio-temporal chaos. (a) Activator control. (b) Inhibitor control. (c) Phase diagram for non-diagonal control (\ref{mixed_c}). Colour denotes a specific final state: spatio-temporal chaos (black), stable mixed (Turing-Hopf) state (pink), coarsening (green), bi-stability between travelling waves and a trivial steady state $E_{0}$ (red) and a uniform non-trivial steady state $E_{1}$ (yellow). Parameter values are $a=0.028$, $b=0.053$, $D_{u}=2\cdot 10^{-5}$, $D_{v}=10^{-5}$.}\label{uvcontrol}
\end{figure*}
where $\Arctan$ corresponds to the principal value of arctan. We are only interested in the branch $n=0$, as for other values of $n$ the corresponding control term in the system (\ref{GS}) will be very large compared to other terms, making it unfeasible for practical purposes.
In Fig.~\ref{stab_bounds}(a) we show this stability boundary, which corresponds to $q=0$, together with stability boundaries when spatially inhomogeneous perturbations are taken into consideration. For any $q>0$, these boundaries lie below the stability boundary for the uniform mode, as we have explained earlier. Figure \ref{stab_bounds}(b) shows the results of computation of the stability boundary in the $(K,\tau)$ plane for inhibitor control. One can note the approximate symmetry between the stability boundaries for activator and inhibitor control up to a reverse of the sign of the control strength. This is due to the fact that the inhibitor control would result in the characteristic equation similar to (\ref{char_eq}), but with the term $(a+b)K[e^{-\lambda\tau}-1]$ being replaced by $-(v_{1}^{2}+a)K[e^{-\lambda\tau}-1]$, where the difference in the absolute values of $(a+b)$ and $(v_{1}^{2}+a)$ is less than 10\%. Finally, we note that it is only possible to stabilize the steady state $E_{1}$ for negative values of the control strength $K$ in the case of activator control, and only for positive $K$ in the case of inhibitor control.

In order to understand the full dynamics of the system (\ref{GS}) with time-delayed feedback control, we solve this system numerically and record the ultimate stage of time evolution. System parameters are fixed and taken to be  $D_{u}=2\cdot 10^{-5}$ and $D_{v}=10^{-5}$, while kinetic parameters are $a=0.028$ and $b=0.053$. In each case we solve system (\ref{GS}) on an interval of length $L=2.5$ using an explicit Euler scheme with a spatial resolution of 256 mesh points, a time-step of $\Delta t=0.05$ and periodic boundary conditions. Initial conditions are chosen to be a random localized perturbation of $E_{0}$ in the middle of the domain which then evolves into a spatio-temporally chaotic state. After the spatio-temporal chaos has developed, control is switched on at $t=3000$ and its effects on the time evolution are studied. Time duration of each run is $6000$ in the regime of spatio-temporal chaos, and $7000$ in the case of control of travelling waves.

\begin{figure*}
\hspace{-1.5cm}\includegraphics[width=17.5cm]{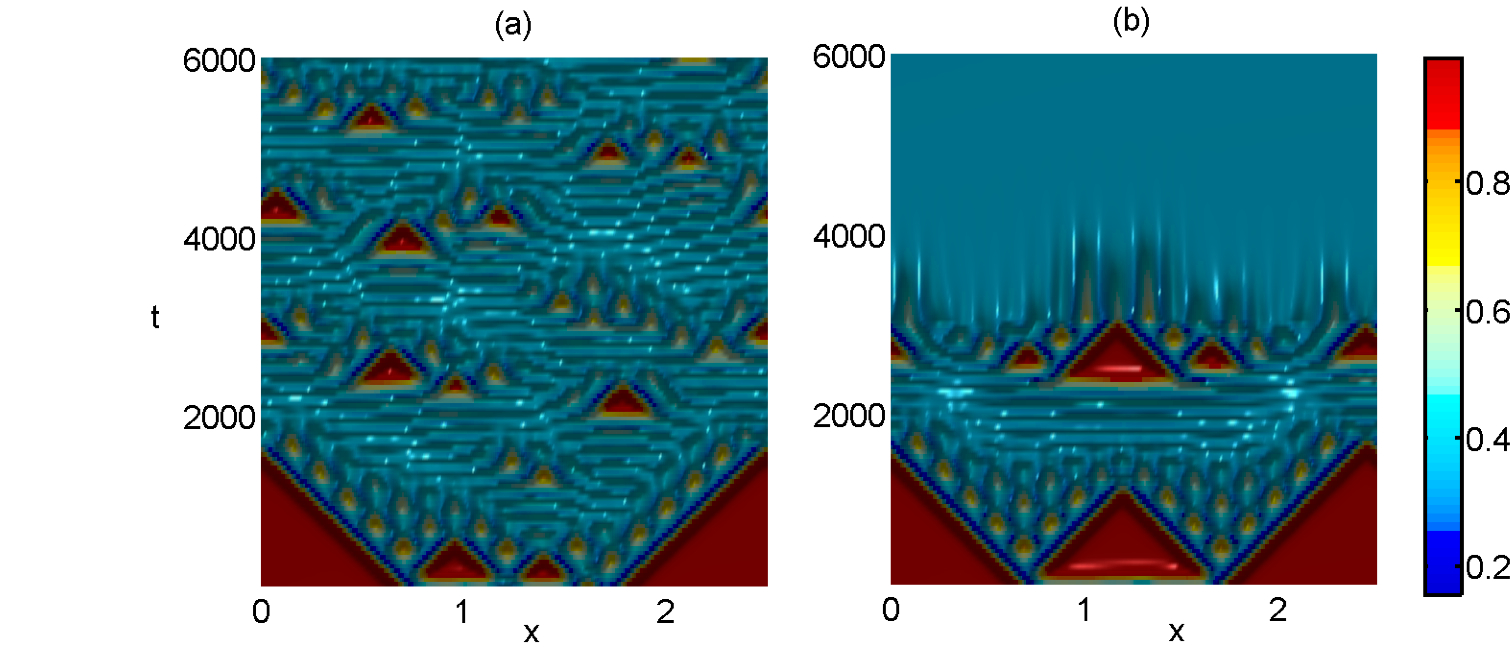}
\caption{(Color online) Space-time plot in the case of activator control. (a) Spatio-temporal chaos. (b) Non-trivial steady state $E_{1}$. The colour code corresponds to the values of $u(x,t)$. Parameter values are $a=0.028$, $b=0.053$, $D_{u}=2\cdot 10^{-5}$, $D_{v}=10^{-5}$, (a) $K=-0.4$, $\tau=0.1$ and (b) $K=-0.8$, $\tau=0.6$. Control is switched on at $t=3000$.}\label{usol}
\end{figure*}

The results of numerical simulations of activator/inhibitor control are shown in Fig. \ref{uvcontrol}(a) and (b). First of all, one can note a characteristic scaling of $|K|\propto 1/\tau$, which is quite natural bearing in mind that if the time delay is small, then the absolute value of the difference $|u(t-\tau)-u(t)|$ (or $|v(t-\tau)-v(t)|$) is small, and therefore a much higher value of the control strength $|K|$ is required to achieve a significant effect on the dynamics. In the case of activator control with $K<0$ 
or inhibitor control with $K>0$, the only possible transition is from a spatio-temporally chaotic state to a non-trivial homogeneous state 
$E_{1}$, as shown in Fig.~\ref{usol}. One should note that when the control strength is not sufficiently large, the system remains in the spatio-temporally chaotic state, as shown in Fig.~\ref{usol}(a). The numerical boundary separating spatio-temporally chaotic state from a stable non-trivial homogeneous state shown in Figs.~\ref{usol}(a) and \ref{usol}(b) coincides with an earlier found analytical expression for this boundary given by Eq. (\ref{stability}).

\begin{figure*}
\hspace{-1cm}\includegraphics[width=18.5cm]{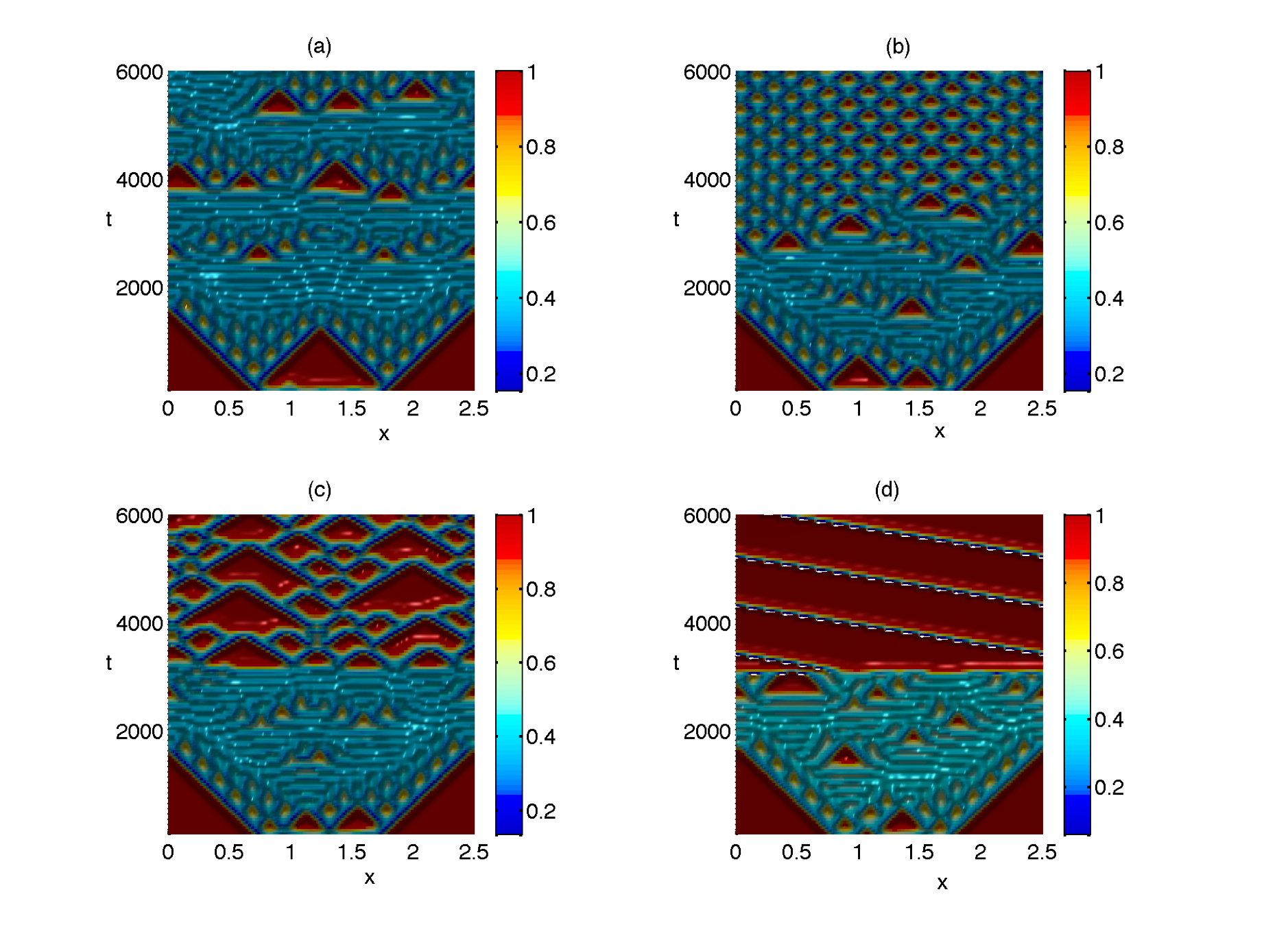}
\caption{(Color online) Space-time plot in the case of inhibitor control. (a) Spatio-temporal chaos. (b) Mixed (Turing-Hopf) mode. (c) Coarsening. (d) Transition to travelling waves. Parameter values are $a=0.028$, $b=0.053$, $D_{u}=2\cdot 10^{-5}$, $D_{v}=10^{-5}$,  (a) $K=-0.05$, $\tau=0.3$, (b) $K=-0.3$, $\tau=0.35$, (c) $K=-0.4$, $\tau=0.4$ and (d) $K=-0.6$, $\tau=0.75$. Control is switched on at $t=3000$.}\label{sol01}
\end{figure*}

In the opposite case, for activator control with $K>0$ or inhibitor control with $K<0$, when the 
control strength $|K|$ and the time delay $\tau$ are sufficiently small, the system remains in the state of spatio-temporal chaos (see Fig.~~\ref{sol01}(a)). For higher values of $|K|\tau$ one observes stable mixed or Turing-Hopf modes shown in Fig.~\ref{sol01}(b). These states are characterized by both spatial and temporal periodicity. Earlier work on co-dimension two Turing-Hopf states suggests that 
generically these are quite stable close to the Turing-Hopf boundary, and away 
from it they turn into either pure Hopf oscillations or Turing structures 
\cite{dWLDB,JBBES,MWBS}. In the case of the Gray-Scott system this does not 
happen as the Turing instability boundary is completely enclosed inside the 
region of Hopf instability.

For yet higher values of $|K|\tau$, isolated regions coalesce, and the system exhibits coarsening, as shown in Fig.~\ref{sol01}(c). 
Further increase in $|K|\tau$ leads to a bistability between a homogeneous trivial state $E_{0}$ or a localized pulse travelling with a 
constant velocity. This bistability here refers to a regime where for the same values of parameters and random initial conditions, the system will evolve either as shown in Fig.~\ref{sol01}(d), i.e. develop into a localized pulse, or it will develop in a manner qualitatively similar to the one shown in Fig.~\ref{usol}(b), except that the evolution leads to a uniform trivial steady state $E_{0}$. We also note that the time-delayed feedback of the form used in (\ref{GS}) is non-invasive only in the case of stabilization of a uniform steady state $E_{1}$. In all other situations, although the feedback term does not vanish, its norm constitutes less than 10\% compared to other terms in the system (\ref{GS}).

Another possibility of control is the so-called diagonal control, which corresponds to the unity coupling matrix $A$.
This type of control has been successfully used to stabilize unstable spatio-temporal breathing in a reaction-diffusion model for the double barrier resonant tunneling diode \cite{UAJS}. Numerical simulations suggest that for the Gray-Scott model this type of control does not qualitatively affect the dynamics as it leaves the system in a spatio-temporally chaotic state for {\it any} values of the control strength $K$ and the time delay $\tau$.

In the case of non-diagonal control that involves both species we introduce a variable phase $\phi$ into the coupling matrix as follows
\begin{equation}\label{mixed_c}
A=\left(
\begin{array}{cc}
\cos\phi&\sin\phi\\
-\sin\phi&\cos\phi
\end{array}
\right).
\end{equation}
Such kind of control occurs, for example, in laser systems, where the optical phase can be used as an additional control parameter \cite{FAYSL,SHWSH}. Figure \ref{uvcontrol}(c) shows that for this type of control, the final state of the system is completely determined by the sign and size of non-diagonal control terms
and shows no dependence on the values of the diagonal coupling terms. It is worth noting that in the case of {\it negative feedback} $(-\pi\leq\phi\leq 0)$, i.e. when the roles of activator and inhibitor in the control term are reversed, one observes a range of possible dynamical regimes, such as stable Turing-Hopf modes, coarsening and bistability between travelling waves and a trivial steady state $E_{0}$.
On the other hand, for {\it positive feedback} with $0\leq\phi\leq\pi$, the only possible transition is from a spatio-temporally chaotic state to a uniform non-trivial steady state $E_{1}$.

Beta and Mikhailov have previously shown for the complex Ginzburg-Landau equation that in the Benjamin-Feir regime, which gives rise to 
spatio-temporal chaos, the global TDFC is unable to stabilize uniform oscillations \cite{Beta2}. Our simulations suggest that a similar conclusion holds for local time-delayed feedback: while it is possible to recover homogeneous states, as well as stable mixed modes, it appears impossible to stabilize uniform oscillations in a reaction-diffusion system.

\section{Control of travelling waves}

In this section we consider a situation when without control the system supports stable travelling waves. By virtue of choosing periodic boundary conditions, if there is more than one pulse in the interval, all pulses are elastically reflected from each other and move without changes through the boundaries.

\begin{figure*}
\hspace{-0.5cm}\includegraphics[width=17.5cm]{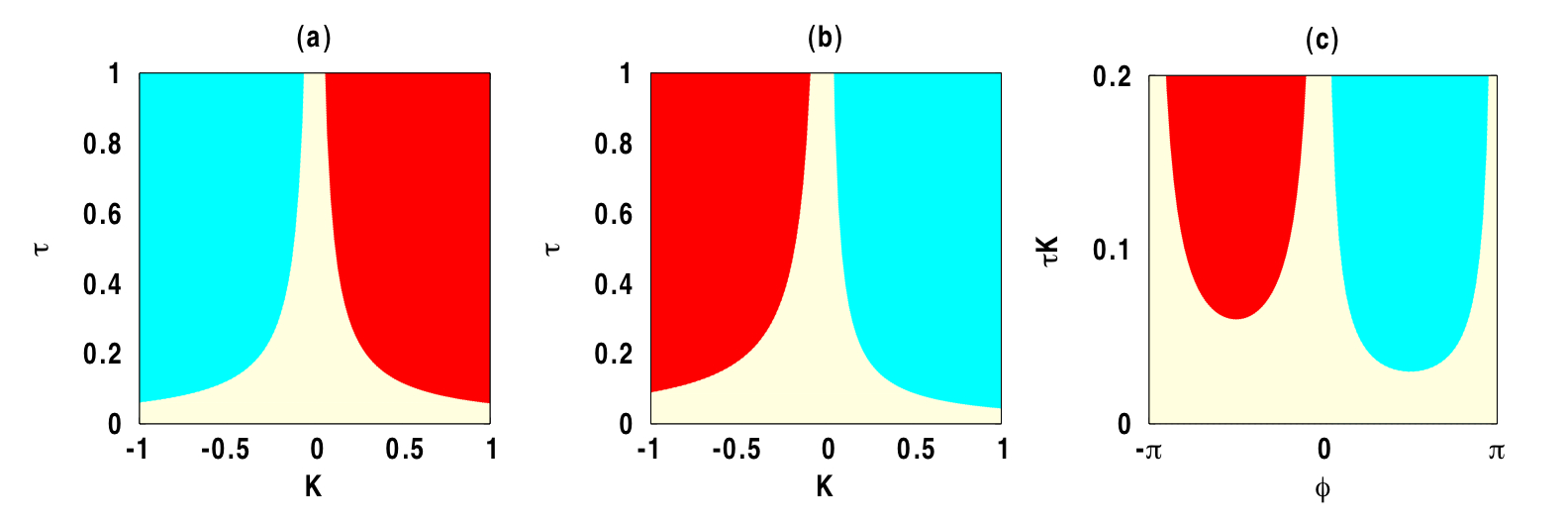}
\caption{(Color online) Phase diagram for control of travelling waves. (a) Activator control. (b) Inhibitor control. (c) Phase diagram for non-diagonal control (\ref{mixed_c}) of travelling waves. Colour denotes a specific final state: stable travelling wave (beige),  stationary Turing pattern (cyan), bi-stability between travelling waves and a trivial steady state $E_{0}$ (red). Parameter values are $a=0.022$, $b=0.053$, $D_{u}=2\cdot 10^{-5}$, $D_{v}=10^{-5}$.}\label{trav}
\end{figure*}

\begin{figure*}
\includegraphics[width=16.5cm]{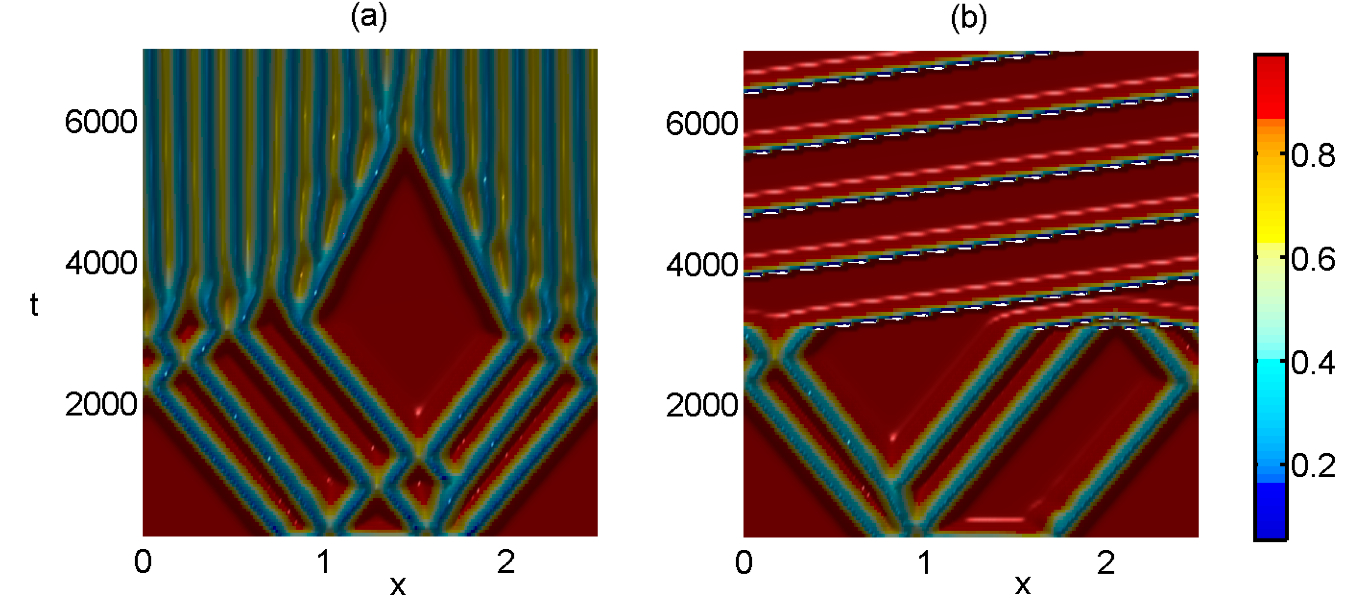}
\caption{(Color online) Space-time plot of control of travelling waves. (a) Activator control: Development of a stationary Turing pattern. (b) Inhibitor control: Transition to travelling wave. Parameter values are $a=0.022$, $b=0.053$, $D_{u}=2\cdot 10^{-5}$, $D_{v}=10^{-5}$, (a) $K=-0.5$, $\tau=0.6$ and (b) $K=-0.8$, $\tau=0.4$. Control is switched on at $t=3000$.}\label{trav_sol}
\end{figure*}

As before, we start our analysis by considering single-species control of the form (\ref{ss_control}), and the results are shown in Fig.~\ref{trav}(a) and (b). In the case of activator control with $K<0$ or inhibitor control with $K>0$, the stable travelling waves are eventually transformed into stable spatially periodic Turing patterns by means of wave-splitting, as shown in Fig.~\ref{trav_sol}(a). As has been noted by Petrov {\it et al.} \cite{PSS}, this happens due to the fact that the medium becomes saturated with waves which are reflected from each other. Numerical simulations suggest that the higher the value of $|K|\tau$, the longer it will take for the wave splitting to result in a stationary spatially periodic pattern and that TDFC can stabilize unstable multi-spike steady states. This provides a mechanism of transformation from a stable propagating pulse to a stable Turing pattern. Finally, we note that the activator ($K>0$) or inhibitor ($K<0$) control  shown in Fig.~\ref{trav}(a) and (b), can only provide the bistability between a homogeneous trivial state $E_{0}$ and travelling waves. For illustration purposes, we show in Figure~\ref{trav_sol}(b) the transition to stable travelling pulses, however in the same parameter regime the system may also evolve into a homogeneous state $E_{0}$.

Figure~\ref{trav}(c) shows the effects of mixed control on the dynamics of travelling waves with some similarity to the earlier picture of mixed control of spatio-temporal chaos (Fig.~\ref{uvcontrol}(c)). There is still almost no dependence on the strength and sign of activator control, but the region of stationary Turing patterns is noticeably larger than that of bistability between a homogeneous trivial steady state $E_{0}$ and a travelling wave. As in the case of spatio-temporal chaos, the diagonal control of travelling waves does not qualitatively change the dynamics, and the system remains in the state of supporting elastic travelling waves.
\begin{figure*}
\hspace{-1.5cm}\includegraphics[width=19cm]{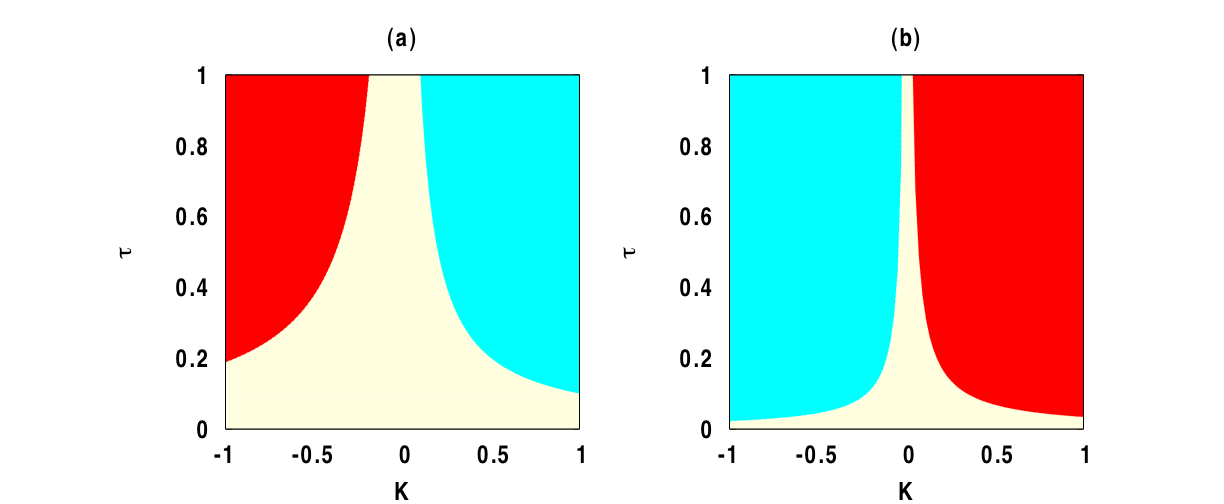}
\caption{(Color online) Phase diagram for control of travelling waves. (a) Activator $uv$-control. (b) Inhibitor $vu$-control. Colour denotes a specific final state: stable travelling wave (beige),  stationary Turing pattern (cyan), bistability between travelling waves and a trivial steady state $E_{0}$ (red). Parameter values are $a=0.022$, $b=0.053$, $D_{u}=2\cdot 10^{-5}$, $D_{v}=10^{-5}$.}\label{travUV}
\end{figure*}

When compared to recent work on TDFC of travelling pulses in the FitzHugh-Nagumo model \cite{DAH08}, our results show both some similarities, and also certain differences. In both cases, the control domain of the inhibitor control is smaller than that of the activator control. It is worth noting that in \cite{DAH08} non-diagonal control was not applied in the form of mixed control (\ref{mixed_c}), involving both activator and inhibitor variables, but was in the form of
\[
A=\left(
\begin{array}{ll}
0&\pm 1\\
0&0
\end{array}
\right)\hspace{0.5cm} (uv-{\rm control}),\hspace{0.5cm}
{\rm or}\hspace{0.5cm}
A=\left(
\begin{array}{ll}
0&0\\
\pm 1&0
\end{array}
\right)\hspace{0.5cm}(vu-{\rm control}).
\]
Suppression of the traveling pulse in case of $uv$-control was only possible for the 
minus sign, while in case of $vu$-control suppression was only found for 
the plus sign, following a strict symmetry with respect to the operation $K \to -K$.
In some sense this is similar to Fig.~\ref{travUV} where the uniform steady state (red) 
can be stabilized only for negative $uv$ feedback or positive $vu$ feedback.

The main difference, however, is that the non-diagonal control of travelling pulses in the Gray-Scott model in the form Eq.~(\ref{mixed_c}) is successful for both signs of the control strength $K$, even though for one sign of $K$ it produces stationary Turing patterns and for another it leads to a homogeneous trivial steady state $E_{0}$. Finally, we note that $vu$-control is able to stabilize the uniform steady state or stationary Turing patterns for significantly smaller values of $K$ and $\tau$ as compared to $uv$-control. Also, Figure~\ref{travUV} shows that the roles of activator and inhibitor are reversed for the same values of $K$ and $\tau$ when compared to Fig.~\ref{trav}.

\section{Conclusions}

In this paper we have studied the effects of TDFC on the dynamics of the Gray-Scott model. When the original system is in the spatio-temporally chaotic state, an activator control with $K<0$ or inhibitor control with $K>0$ transforms it into a homogeneous non-trivial steady state $E_{1}$. A much wider range of possible states can be reached for an activator control with $K>0$ or inhibitor control with $K<0$, and these include the stable mixed Turing-Hopf modes, coarsening, as well as a regime of bistability between a stable pulse and a homogeneous trivial state $E_{0}$. Diagonal control appears to be ineffective as it cannot control the spatio-temporal chaos. For mixed type of TDFC, the simulations suggest that the sign of non-diagonal terms in the coupling matrix plays a crucial role in determining the final state of the system. As in the case of global TDFC, the local control is unable to stabilize uniform periodic oscillations.

When the TDFC is applied to travelling pulses of the Gray-Scott model, there are two possibilities for the final state. Activator control with $K<0$ or inhibitor control with $K>0$ transforms the system into a regime of wave splitting, and after the medium is saturated with the waves, it settles on a stationary spatially periodic Turing pattern. The higher is the value of $|K|\tau$, the longer it takes for the system to reach such a pattern. On the other hand, activator control with $K>0$ or inhibitor control with $K<0$ suppresses spatial inhomogeneities and provides bistability between a stable travelling pulse and a homogeneous trivial state, similar to the control of spatio-temporal chaos.

The control schemes studied in this paper can be used for control of spatio-temporal chaos in other spatially extended systems. While stabilization of uniform periodic oscillations may remain elusive, local TDFC can provide a variety of interesting dynamical regimes, from stationary patterns to mixed modes and travelling localized pulses. Numerical simulations suggest that in many cases the control strength does not have to be very high, provided time-delay is large enough. This interplay between the control strength and time delay should prove beneficial for the development of experimental implementation of specific control schemes.

\section*{Acknowledgements} Y.K. acknowledges the support from the EPSRC via EPSRC Postdoctoral Fellowship (Grant EP/E045073/1). This work was partially supported by Deutsche Forschungsgemeinschaft in the framework of Sfb 555.

\end{document}